\documentclass[aps,pra,twocolumn,superscriptaddress,nofootinbib]{revtex4-1}

\usepackage{dirtytalk}
\usepackage{bm}
\usepackage{amsmath,amssymb}
\usepackage{amsfonts}
\usepackage{layouts}
\usepackage{mathtools}
\usepackage{dsfont}
\usepackage{graphicx}
\usepackage{float}
\usepackage{subfigure} 
\usepackage{verbatim}
\usepackage{amsfonts}
\usepackage{bbold}
\usepackage{dcolumn}
\usepackage{bm}
\usepackage{color}
\usepackage[dvipsnames]{xcolor}
\usepackage{listings}
\definecolor{blueprl}{RGB}{46,48,146}
\usepackage{mathrsfs}
\usepackage{filecontents}
\usepackage{times} %font times new roman
\newcommand{\bra}[1]{\mbox{$\langle #1 |$}}
\newcommand{\ket}[1]{\mbox{$| #1 \rangle$}}
\def\ketbra#1#2{{\vert#1\rangle\!\langle#2\vert}}
\newcommand{\braket}[2]{\mbox{$\langle #1 | #2 \rangle$}}
\usepackage{lipsum} % to create dummy text
\newcommand{\xdownarrow}[1]{%
  {\left\downarrow\vbox to #1{}\right.\kern-\nulldelimiterspace}
}

\makeatletter
\newcommand*{\balancecolsandclearpage}{%
  \close@column@grid
  \clearpage
  \twocolumngrid
}
\makeatother

\usepackage{gensymb}
\usepackage[breaklinks=true]{hyperref}
\hypersetup{
colorlinks   = true, %Colours links instead of ugly boxes
urlcolor     = blue, %Colour for external hyperlinks
linkcolor    = blue, %Colour of internal links
citecolor    = red %Colour of citations
}
\usepackage{comment}
\usepackage[normalem]{ulem}
\usepackage{xcolor}
\usepackage{graphicx}
\usepackage{cleveref}
\crefname{equation}{Eq.}{Eqs.}
\Crefname{equation}{Equation}{Equations}
\crefname{figure}{Fig.}{Figs.}
\Crefname{figure}{Figure}{Figures}
\crefname{figure}{Fig.}{Figs.}
\Crefname{figure}{Figure}{Figures}
\crefname{section}{Sec.}{Secs.}
\Crefname{section}{Section}{Sections}
\crefname{appendix}{Appendix}{Appendices}
\Crefname{appendix}{Appendix}{Appendices}
\crefname{table}{Table}{Tables}
\Crefname{table}{Table}{Tables}

\newcommand{\Id}{\mathbb{1}}

\newcommand{\oneinjectedscissor}{1\text{-scissor}}
\newcommand{\twoinjectedscissor}{2\text{-scissor}}
\newcommand{\oneinjectedscissors}{1\text{-scissors}}
\newcommand{\threeinjectedscissor}{3\text{-scissor}}

\newcommand{\seveninjectedscissor}{7\text{-scissor}}

\newcommand{\Nscissors}{N\text{ 1-scissors}}
\newcommand{\twoscissors}{\text{two 1-scissors}}
\newcommand{\threescissors}{\text{three 1-scissors}}
\newcommand{\twocatrelay}{2\text{-cat tele-amplification}}

\newcommand{\fourcatrelay}{4\text{-cat tele-amplification}}

\newcommand{\smallfrac}[2]{{\mbox{$\frac{#1}{#2}$}}}

\makeatletter
\def\@bibdataout@aps{%
 \immediate\write\@bibdataout{%
  @CONTROL{%
   apsrev41Control,author="08",editor="1",pages="0",title="0",year="1",eprint="1"%
  }%
 }%
 \if@filesw
  \immediate\write\@auxout{\string\citation{apsrev41Control}}%
 \fi
}%
\makeatother % Phew.
\begin{document}

\title{
Generalized quantum scissors for noiseless linear amplification
}

\author{Matthew S. Winnel}\email{matthew.winnel@uqconnect.edu.au}
\affiliation{Centre for Quantum Computation and Communication Technology, School of Mathematics and Physics, University of Queensland, St Lucia, Queensland 4072, Australia}
\author{Nedasadat Hosseinidehaj} 
\affiliation{Centre for Quantum Computation and Communication Technology, School of Mathematics and Physics, University of Queensland, St Lucia, Queensland 4072, Australia}
\author{Timothy C. Ralph}
\affiliation{Centre for Quantum Computation and Communication Technology, School of Mathematics and Physics, University of Queensland, St Lucia, Queensland 4072, Australia}

\date{\today}

\begin{abstract}
We generalize the concept of optical state truncation and noiseless linear amplification to enable truncation of the Fock-state expansion of an optical state to higher order and to simultaneously amplify it using linear optics. The resulting generalized quantum scissors are more efficient for noiseless linear amplification than employing multiple scissors in parallel and are experimentally practical. As a particular example, we focus on a third-order scissor device and demonstrate advantages in terms of fidelity with the target state, probability of success, distillable entanglement, and the amount of non-Gaussianity introduced.
\end{abstract}

\maketitle

%====================================================
%====================================================
%====================================================
%====================================================

%==============
%====================================================
%\section{\label{sec:intro}Introduction}
%====================================================

The no-cloning theorem~\cite{wootters1982single} forbids the deterministic, linear (i.e. phase insensitive) amplification of a quantum state. Hence, all deterministic linear amplifiers must introduce noise~\cite{caves1981quantum-mechanical}. Nevertheless, non-deterministic noiseless linear amplification is possible if the amplifier is allowed to operate in a probabilistic, but heralded way, and the alphabet of states has an energy bound \cite{ralph2009nondeterministic,xiang2010heralded}.

Noiseless linear amplification has proven a very useful technique in quantum optics with numerous experimental demonstrations \cite{barbieri2011nondeterministic} such as distillation of entanglement \cite{xiang2010heralded, PhysRevA.100.022315}, purification of entanglement \cite{Ulanov15}, amplification of qubits \cite{Kosis13} and enhanced metrology \cite{Usuga10}. Proposed applications include continuous variable error correction \cite{Ralph11}, quantum key distribution~\cite{Blandino2012CV-QKD, Ghalaii_2020, ghalaii2020discretemodulation} and quantum repeaters~\cite{dias2018quantum, PhysRevResearch.2.013310}, and discrete variable Bell inequalities~\cite{osorio2012heralded}. 

The non-deterministic quantum scissor introduced by Pegg et al.~\cite{pegg1998optical} truncates an input optical field to first order, retaining only the vacuum and one-photon components of the input state. %generalizations of the quantum scissor to higher order have been made~\cite{koniorczyk2000general,villas-boas2001recurrence,miranowicz2005optical-state}.
In their original proposal, Ralph and Lund introduced a modified scissor device which truncates an input state to first order and simultaneously amplifies it by increasing the amplitude of the one-photon component relative to the vacuum component~\cite{ralph2009nondeterministic}. For input states of small amplitude, the modified scissor acts as an ideal noiseless linear amplifier (NLA). 

In order to go beyond small input amplitudes, Ralph and Lund proposed employing multiple quantum scissors in parallel~\cite{ralph2009nondeterministic}. For a large finite number of scissors, the set-up acts as an ideal NLA, however with a vanishing probability of success. If a small number of scissors are used, the amplification is ``distorted'', that is, the Fock coefficients of the state obtained are multiplied by different constants than those required for ideal linear amplification. Other methods for NLA do not truncate the state but still distort Fock components higher than one~\cite{zavatta2010high-fidelity,Ulanov15}.

To improve the one photon scissor-based NLA, there have been attempts to generalize the modified quantum scissor to two photons~\cite{jeffers2010nondeterministic}, however the resulting device imposes an undesired non-linear sign change to the two photon component of the amplified state. A generalization of the original quantum scissor to higher order has also been made but this does not allow for amplification~\cite{koniorczyk2000general,
villas-boas2001recurrence,miranowicz2005optical-state}. Alternatively, a protocol that could truncate and amplify without distortion was described in Ref.~\cite{McMahon_2014}, however this solution is impractical as it requires a massive high order optical non-linearity. %, however, the Fock coefficients of the state obtained are multiplied by different constants than those required for ideal truncation and amplification. 
As a result, no demonstration of NLA without distortion of the higher order Fock state components has been achieved, thus seriously limiting future applications.

%Many protocols and experiments for quantum teleportation, entanglement distillation, and key distribution employ quantum scissors. For instance, scissors have been used in CV error correction for CV repeaters~\cite{dias2018quantum}, entanglement distillation~\cite{seshadreesan2019continuous-variable}, and  QKD~\cite{ghalaii2018longdistance,ghalaii2019discretemodulation}. They have also appeared in experimental proposals~\cite{zdemir2001quantum} and implementations~\cite{hu2019entanglement,bruno2013complete,xiang2010heralded,ferreyrol2010implementation}, see in particular Ref.~\cite{barbieri2011nondeterministic} for a review, and they have also seen applications in discrete variables~\cite{osorio2012heralded}.

%It is worth noting that there are other ways to do noiseless linear amplification, such as coherent combination of photon ad and subtraction~\cite{fiur2009engineering,zavatta2010high-fidelity}, and Ref.~\cite{McMahon_2014} which introduces an optimal system for a nondeterministic NLA. The advantage of a scissors-based NLA is that the gain may be easily tuned by altering the transmission coefficients of the beamsplitters, and scissors can be implemented with only linear optics.

%In all attempts to date, NLA based on higher order quantum scissors result in ``distorted amplification.'' 

Here, we generalize the concept of optical state truncation and amplification and propose a practical linear optical device which can correctly amplify the input state up to higher order. %Our generalized scissors function in a way that is analogous to the original modified quantum scissor and uses only linear optics. 
The device naturally performs noiseless linear amplification without distorting the amplified Fock coefficients.

{\it Generalized scissors}: Suppose the input state in the Fock basis is $\ket{\psi}{=}\sum_{n=0}^\infty c_n |n\rangle.$ An ideal NLA performs the transformation $\hat{T}_\text{ideal}\ket{\psi}{\to}\sum_{n=0}^\infty g^n c_n |n\rangle$, where g is the gain of the NLA. The success probability is zero for any device that achieves this transformation perfectly \cite{Pandey_2013}.

\begin{figure}
              \makebox[\linewidth][c]{\includegraphics[width=1.25\linewidth]{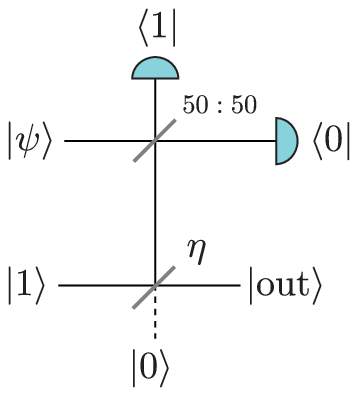}}
          \caption{The original modified single photon quantum scissor ($\oneinjectedscissor$). It consists of the injection of a single photon, and at the two measurement ports, detection of a single photon at one and no photon at the other. The beamsplitter transmissivity $\eta$ sets the gain of the NLA.}
          \label{fig:scissor}
          \end{figure}

The modified single photon quantum scissors~\cite{pegg1998optical, ralph2009nondeterministic} truncate and amplify an optical state in Fock space and is shown in \cref{fig:scissor}. It performs the transformation
\begin{align}\label{eq:1scissor}
\hat{T}_1 \ket{\psi} &= \sqrt{\frac{1}{2(g^2+1)}} (c_0 |0\rangle \pm g c_1 |1\rangle),
\end{align}
where the gain is $g{=}\sqrt{\eta/(1{-}\eta)}$. The plus sign corresponds to the measurement outcome shown in \cref{fig:scissor}, i.e., detection of a single photon at the upper port and no photons at the port on the right. The minus sign corresponds to the reverse measurement outcome, i.e., no photon at the top and a single photon at the side.%\footnote{The gain is $g=\sqrt{(1-\eta)/\eta}$ if the injected photon comes from the bottom not the side in \cref{fig:scissor}}

This device is called a quantum scissor since all Fock components greater than one are truncated. For the device to operate as an ideal NLA, the two photon component must be negligible, that is $|g^2c_2|{\ll}|gc_1|$. The one photon scissor acts as an ideal NLA only for small input states amd the effect of the truncation is severe for large input states.

\begin{figure}
      \makebox[\linewidth][c]{\includegraphics[width=0.9\linewidth]{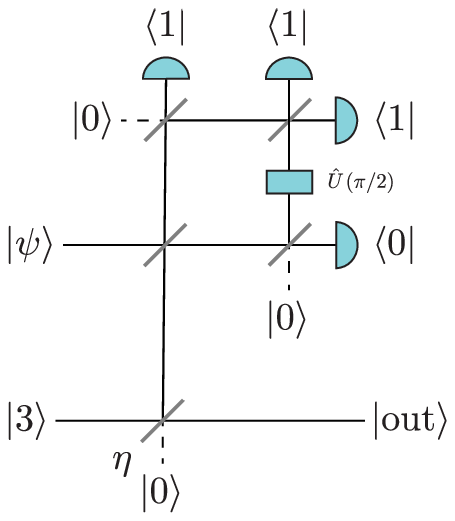}}
          \caption{Our generalized three photon quantum scissor ($\threeinjectedscissor$) with the injection of a three photon state and four-port single photon measurement scheme, with three ports detecting single photons and the remaining port detecting no photons. The upper four beamsplitters are all set to have transmissivity $1/2$. The lower beamsplitter has transmissivity $\eta$ which sets the gain of the NLA. There is a vital $\pi/2$ phase shift on one of the paths.}
          \label{fig:3scissor}
\end{figure}

Our generalized three photon quantum scissor is shown in \cref{fig:3scissor}. It can amplify input states of larger amplitude. It performs the transformation
\begin{multline}\label{eq:3scissor}
\hat{T}_3 \ket{\psi} = \frac{\sqrt{6}}{8} \left(\frac{1}{g^2+1}\right)^\frac{3}{2} \\ \times (c_0 |0\rangle + g c_1 |1\rangle + g^2 c_2 |2\rangle + g^3 c_3 |3\rangle),
\end{multline}
where the gain is $g{=}\sqrt{\eta/(1{-}\eta)}$ and the success probability is
\begin{align}\label{eq:1scissor}
P_3 &= 4 \times | \bra{\psi }\hat{T}^\dagger_3 \hat{T}_3 \ket{\psi} |^2.
\end{align}
Three photons enter the device as a resource,  three single photon detectors register single photons, and one detector registers no photon. Specifically, we consider the measurement outcome  shown in \cref{fig:3scissor}, i.e. detecting three single photons at the top of the device and no photons at the other mode. There are three other possible patterns for getting three ``clicks'' and one ``no click.'' These other click patterns lead to a heralded passive phase shift of the state, but the magnitudes of the Fock components are unchanged. Thus, the success probability is effectively increased by a factor of four.

The three photon scissor truncates and ideally amplifies the input state to third order. The device operates as an ideal NLA as long as $|g^4c_4|{\ll}|g^3c_3|$, which is a major improvement from the single photon scissor, at the cost of a reduced probability of success.

We have found that our device generalizes (at least) to $2^S{-}1$, where $S{=}1,2,3,$ etc.% The next quantum scissor we have found is a seventh order device.% and is presented in \cref{sec:appendix7-scissor}.
%Ref.~\cite{ralph2009nondeterministic} proposed a method for probabilistic amplification using an arrangement of multiple single photon scissors in parallel. However, NLA based on multiple one photon scissors perform distorted amplification since the Fock components have different coefficients than they should for ideal amplification. Only in the limit of a large number of single photon scissors does this set-up approach an ideal NLA. In contrast, our generalized scissors do not suffer from distorted amplification.
%, 
%\begin{equation}
%\text{an equation}
%\end{equation}
For the rest of the paper we will refer to the scissors as $\oneinjectedscissor$, $\threeinjectedscissor$, $\seveninjectedscissor$, etc. The prefix in our notation is the number of photons entering the device as a resource, the number of detectors that register single photons, and the Fock state truncation of the output state. See~\cref{sec:appendixscissors} for derivations of our generalized quantum scissors.

% We present derivations of the scissors in \cref{sec:appendix1-scissor,sec:appendix3-scissor}. 

%====================================================

%\subsection{Two photon scissor}

%====================================================
%====================================================
%====================================================
%====================================================
%====================================================

%====================================================
%\section{Truncation and amplification of coherent states\label{sec:coherent}}
%====================================================

%The aim of this section is to compare our generalized quantum scissors with NLA devices based on multiple $\oneinjectedscissors$. We take the input state to be a coherent state $\ket{\gamma}$.

{\it Amplification of coherent states}: The fidelity $F$ of the output state with the ideally amplified state (i.e. the target state) is useful as a measure of how well the input state has been amplified. When considering the performance of probabilistic amplifiers, it is also important to consider the success probability.% The performance of the devices introduced are compared in terms of fidelities and success probabilities.

\begin{figure*}
\centering
\includegraphics[width=1\textwidth]{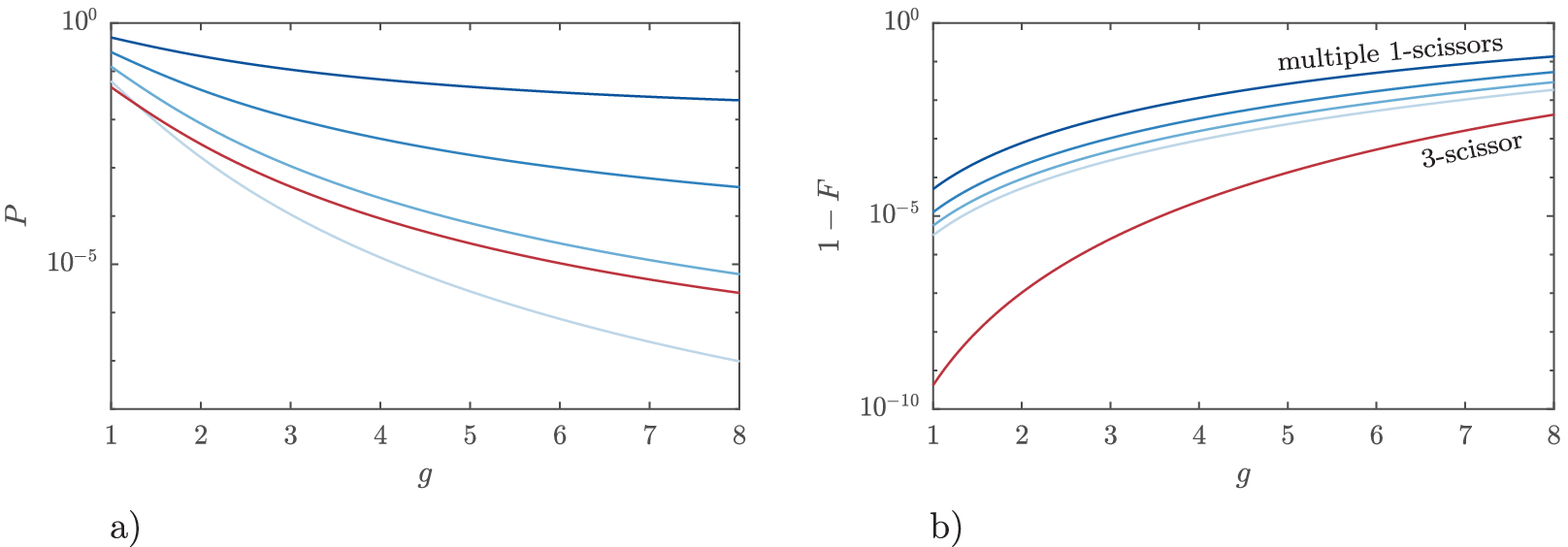}
\caption{a) Probability of success $P$ and b) infidelity $1-F$ versus gain $g$ comparing multiple $\oneinjectedscissors$ in an NLA with the $\threeinjectedscissor$. The input state is a coherent state $\ket{\gamma}$ with magnitude $\gamma=0.1$. The number of $\oneinjectedscissors$ in the NLA is $N=1$ (darkest blue), $N=2$, $N=3$, and $N=4$ (lightest blue). There is a trade-off between fidelity and success probability, however, the $\threeinjectedscissor$ outperforms  the $N=3$ device both in terms of fidelity and success probability.\label{fig:gammapoint1}}
\end{figure*}

In \cref{fig:gammapoint1} we plot the success probability and infidelity $(1{-}F)$ of the output state with the target state $\ket{g\gamma}$ as a function of the gain $g$ for a fixed input coherent state with amplitude $\gamma = 0.1$, and compare generalized scissors with NLA based on multiple $\oneinjectedscissors$ in parallel (see~\cref{sec:appendixPsuc,sec:appendixmultiplescissors} for details). Despite the trade-off between fidelity and success probability we see that our 3-scissor simultaneously achieves higher fidelity \textit{and} success probability than the NLA based on four $\oneinjectedscissors$.

%We write down success probabilities in~\cref{sec:appendixPsuc}.

%We find that the success probability for our $\threeinjectedscissor$ is comparable with the NLA based on three $\oneinjectedscissors$, and the fidelity of our $\threeinjectedscissor$ is greater than that of all practical $N$ devices for moderately small $\gamma$ and $g$. For small $\gamma$, the $\threeinjectedscissor$ is working as an ideal NLA with the infidelity approaching zero.

%\footnote{The success probability of the $\threeinjectedscissor$ shown in the plot is as given in \cref{eq:PROB}, that is, four times the success probability of the outcome of the measurement scheme as in \cref{fig:3scissor}. This is because it is then a fair comparison with the $\oneinjectedscissor$ and multiple $\oneinjectedscissors$, for which it is usually assumed that the phase change on the single photon component is corrected by feeding forward.}

%====================================================
%====================================================
%====================================================%====================================================
%====================================================

%====================================================
%\section{Entanglement distillation with generalized quantum scissors\label{sec:ent}}
%====================================================
%In this section, we compare the $\oneinjectedscissor$ and our $\threeinjectedscissor$ in terms of distillable entanglement and amount of non-Gaussianity introduced.

{\it Entanglement distillation with generalized quantum scissors}: A two mode EPR state with squeezing parameter $r$ is $\ket{\chi} {=}  \sqrt{1{-}\chi^2} \sum_{n=0}^\infty \chi^n \ket{nn}$, where $\chi{=}\tanh{r}$, and the mean photon number is $\bar{n}{=}\sinh^2{r}$. The notation $\ket{nn}$ is shorthand for the two mode Fock state $\ket{n}{\otimes}\ket{n}$.

Placing a $\oneinjectedscissor$ on one arm transforms the EPR state according to
\begin{align}
|\chi\rangle &\to  \sqrt{\frac{1-\chi^2}{2(g^2+1)}} ( \ket{00} + g \chi \ket{11} ).
\end{align}

Placing the $\threeinjectedscissor$ on one arm performs the transformation
\begin{align}
|\chi\rangle &\to   \frac{\sqrt{6}}{8} \sqrt{\frac{1-\chi^2}{(g^2+1)^3} } \sum_{n=0}^3 (g \chi)^n \ket{nn}.
\end{align}

Both scissors herald states that have the form of a truncated EPR state, but with an effective increase in $\chi\to g\chi$. Therefore, the scissors are useful to distill entanglement. These protocols generalize to EPR states distributed through loss, allowing purification of entanglement distributed over long distances~\cite{Ralph11,Ulanov15}.

For such protocols, the $\oneinjectedscissor$ usually works best %in the high-fidelity regime, and is therefore
when limited to small $\chi$ and large loss \cite{dias2018quantum}. The $\threeinjectedscissor$ allows distillation protocols to operate in regimes of higher squeezing and less loss, at the cost of a reduced probability of success, and it also introduces less non-Gaussianity. To demonstrate this, we calculate the entanglement of formation and reverse coherent information~\cite{PhysRevLett.102.050503} in the following. %sections.

%\subsection{Entanglement of formation}

After transmission of one mode of an EPR state through a pure loss channel of transmissivity $T$ followed by either a $\oneinjectedscissor$ or a $\threeinjectedscissor$, we calculate the Gaussian entanglement of formation (GEOF) as an entanglement measure to evaluate the performance of each scissor. The GEOF quantifies the amount of two-mode squeezing required to prepare an entangled state from a classical state~\cite{PhysRevA.96.062338}, which is an upper bound on the entanglement of formation.

\begin{figure}
\includegraphics[width=1\linewidth]{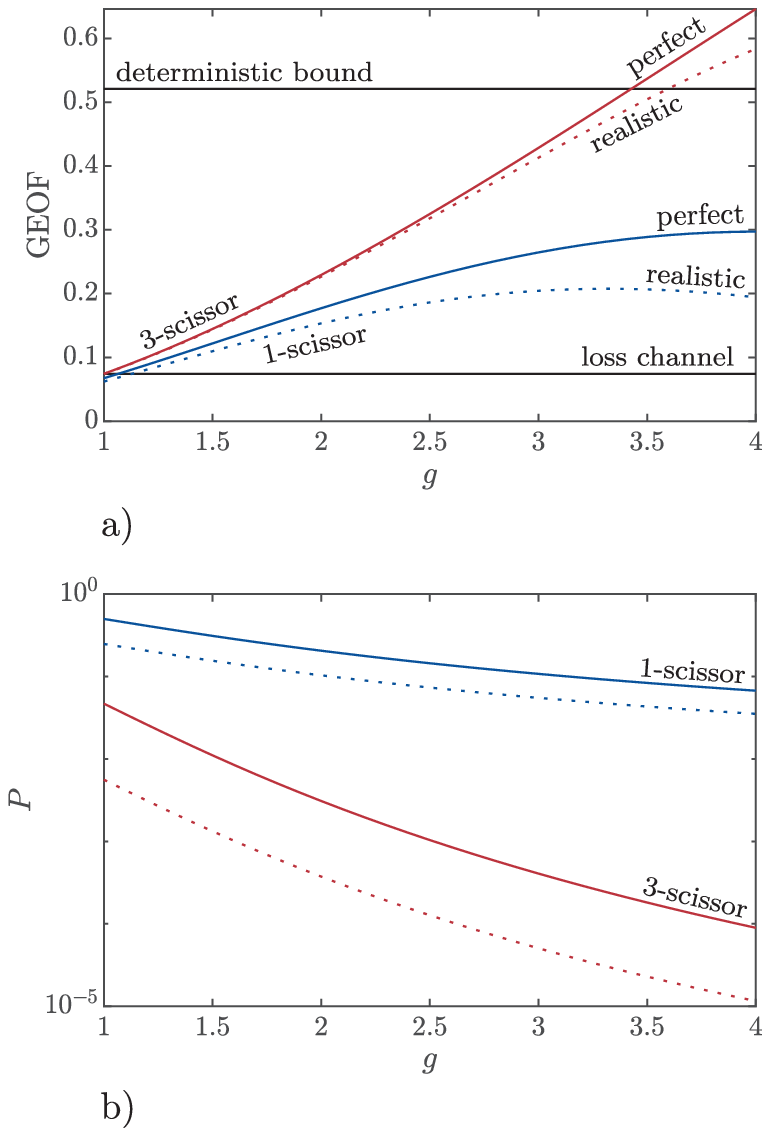}
\caption{a) Gaussian entanglement of formation (GEOF) and b) probability of success $P$ as a function of gain $g$ for an EPR state, with one arm propagated through a lossy channel followed by a $\oneinjectedscissor$ or a $\threeinjectedscissor$ for perfect set-up, and for non-ideal realistic resource and detectors $(\tau_\text{s} = \tau_\text{d} = 0.7)$. Channel transmissivity is $T=0.1$ and EPR parameter is $\chi=0.3$. The ``loss channel'' is GEOF calculated for direct transmission, i.e. no quantum scissor. The ``deterministic bound'' is the amount of entanglement given that an infinitely squeezed state has been sent through the channel~\cite{PhysRevA.96.062338}. This plot shows a situation for which the crossing point of the deterministic bound, the minimum requirement for error correction \cite{dias2018quantum}, can be reached by the $\threeinjectedscissor$, even for a realistic experimental set-up, but is unobtainable by the $\oneinjectedscissor$. Given \cref{fig:gammapoint1} has already demonstrated that the three photon scissor has an improved fidelity and success probability over three or four scissors in parallel, we do not deem it necessary to further investigate entanglement distillation using scissors in parallel.\label{fig:GEOF}}
\end{figure}

The GEOF is calculated following Ref.~\cite{PhysRevA.96.062338} and using results from Ref.~\cite{PhysRevLett.84.2722,PhysRevLett.84.2726}. \Cref{fig:GEOF} shows the GEOF as a function of $g$ for the $\oneinjectedscissor$ and our $\threeinjectedscissor$ given EPR parameter $\chi = 0.3$ and channel transmissivity $T=0.1$. Also shown is the amount of entanglement for the same EPR state and loss channel but with no quantum scissor. The deterministic bound assumes an infinitely squeezed EPR state sent through the same loss channel and no quantum scissor. Crossing the deterministic bound is a necessary condition for the distillation to be useful in error correction or repeater protocols \cite{SpyrosandJosephine}.
The $\threeinjectedscissor$ has a higher GEOF than the $\oneinjectedscissor$ for the same gain. In particular, for these parameters the $\threeinjectedscissor$ crosses the deterministic bound whilst the $\oneinjectedscissor$ is unable to cross this bound.

%Inefficient detectors reduces the success probability but has little effect on the GEOF. Loss at the resource scrambles the correlations for the $\threeinjectedscissor$, however, the GEOF is still strong. For the $\oneinjectedscissor$, loss events at the resource prevents the device from doing anything useful (the device only outputs vacuum), so the entanglement decreases more rapidly with an inefficient resource. These results show that the $\threeinjectedscissor$ performs well despite realistic considerations.

%\begin{figure}
%\includegraphics[width=1\linewidth]{./GEOF.eps}
%\caption{\label{fig:GEOF}}
%\end{figure}

%\subsection{The non-Gaussian effect of scissors}

To demonstrate the non-Gaussian effect of the scissors, we calculate the total reverse coherent information (RCI)~\cite{PhysRevLett.102.050503}, and compare it with the the Gaussian RCI, i.e. the total RCI calculated for a Gaussian state with the same covariance matrix. The RCI gives a lower bound on the distillable entanglement~\cite{PhysRevLett.102.210501} as outlined in \cref{sec:appendixRCI}.

\begin{figure}
\includegraphics[width=1\linewidth]{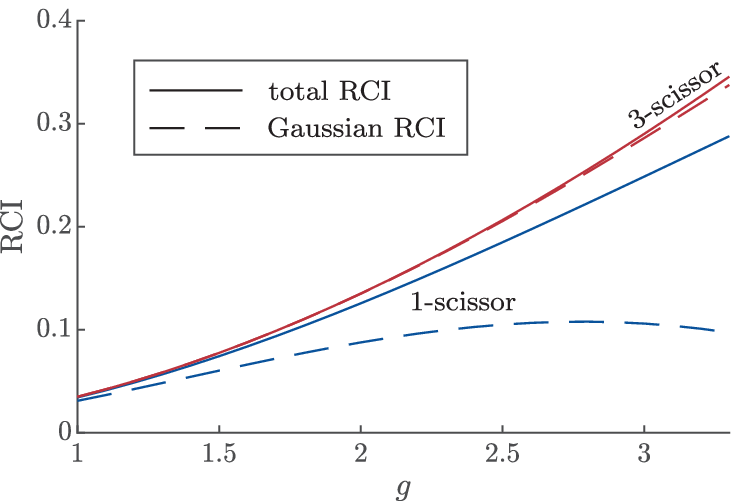}
\caption{Total reverse coherent information (RCI) and Gaussian RCI as a function of gain $g$ for an EPR state, with one arm propagated through a lossy channel followed by a $\oneinjectedscissor$ or a $\threeinjectedscissor$. Channel transmissivity is $T=0.1$ and EPR parameter is $\chi=0.3$. Our $\threeinjectedscissor$ introduces less non-Gaussianity than the $\oneinjectedscissor$.\label{fig:RCI}}
\end{figure}

We plot the total RCI and Gaussian RCI as a function of gain in \cref{fig:RCI} for EPR parameter $\chi=0.3$ and channel transmissivity $T=0.1$. This plot demonstrates that for these parameters, the non-Gaussian entangled state heralded after the $\oneinjectedscissor$ suffers severely from unwanted non-Gaussianity, whereas, the non-Gaussianity introduced by the $\threeinjectedscissor$ is not so harsh.

%These values of $\chi$ and $T$ were chosen as to clearly demonstrate that the $\threeinjectedscissor$ introduces less non-Gaussianity and noise due to the truncation for relatively large input states compared with the $\oneinjectedscissor$.

{\it Imperfect operations}: An important consideration is how the performance of the 3-scissor is affected by experimental imperfections. %We consider the impact of non-perfect detectors on the fidelity. 
Single photon detectors with quantum efficiency $\tau_\text{d}$ can be modelled by a lossy channel with transmissivity $\tau_\text{d}$ followed by a perfect single photon detector. We find that non-perfect efficiency impacts the success probability but has a small impact on the fidelity, see~\cref{sec:appendixDetectors} for details and more plots. %This can be explained since losing photons before the detectors will increase the failure rate of the device, but when the device does succeed, it is very unlikely (at least in this regime) that the correct measurement outcome was obtained during loss events- in other words, for small input states there is approximately 3 photons in the device and recall that we require three clicks and one no-click, so if any losses occur, we must have had at least 4 photons in the device to begin with, and the loss must have happened in the vacuum detection mode while the other three photons must be in exactly in the other three modes (sounds pretty unlikely).
Typically it is more feasible in an experiment to use on-off photon detectors. On-off detectors cannot discriminate between different numbers of photons, but can only distinguish between vacuum and non-vacuum. Again, for input states with small amplitudes, %scissor work approximately well with 
the effect of on-off detection on the fidelity is small.%, see \cref{sec:appendixDetectors}.

The reason that our generalized scissors are robust to these practical issues is due to the detection scheme. The detection scheme separates all the light into several modes and performs single photon detection on those modes. In the high fidelity regime, there is only a small number of photons in the device at any one time (if there was not, the device would not be working in a high fidelity regime), and so errors due to on-off detection or imperfect detectors are rare.

Another important practical aspect of the device is the resource mode. The efficiency of the resource is modelled by a lossy channel with transmissivity $\tau_\text{s}$ following preparation of the Fock state resource, $|1\rangle$ for the $\oneinjectedscissor$ and $|3\rangle$ for the $\threeinjectedscissor$. The success probability and the fidelity are both negatively impacted, but for coherent states, not severely. For coherent state inputs, the $\threeinjectedscissor$ still performs ideal truncation and amplification up to two photons if two photons rather than three are injected into the device, of course with a reduced fidelity since the truncation is at second order not three. This surprising result allows the $\threeinjectedscissor$ to keep performing well even with loss on the resource mode. This result generalizes to all scissors, i.e., the $\seveninjectedscissor$ ideally truncates and amplifies coherent states even if less than seven photons are injected into the device.

In \cref{fig:GEOF} we also include the effect of inefficient detectors and an inefficient resource ($\tau_\text{s}=\tau_\text{d}=0.7$). We find that under realistic conditions the 3-scissor can still distill entanglement above the deterministic bound under conditions for which this is impossible for the 1-scissor.

{\it Discussion and conclusion}: %Despite generalizations of the quantum scissor to order greater than one having been made with some success~\cite{jeffers2010nondeterministic,koniorczyk2000general,villas-boas2001recurrence,miranowicz2005optical-state}, no demonstration of higher order scissor-based NLA without distortion of the Fock coefficients has been introduced previous to our device.
We have shown our generalized scissors %ideally 
amplify and truncate arbitrary input states without distorting the Fock coefficients, assuming perfect implementation. Considering more realistic devices, we find that in the working regime of high fidelity, imperfect single photon detectors, or employing on-off detectors %in place of single photon detectors which is experimentally more feasible, 
has little effect on the fidelity and only impacts the success probability. Of greater importance %though 
is the efficiency of the Fock state resource. For coherent state inputs, we find that the $\threeinjectedscissor$ device is naturally and surprisingly robust to non-ideal resource efficiency.  %, due to the ``$\twoinjectedscissor$'' result as outlined in \cref{sec:appendix2-scissor}.
Realistic devices perform well for entanglement distillation as well.

Another possible application of our scissors would be to engineer optical states~\cite{escher2005synthesis}. For example, making a slight change to our $\threeinjectedscissor$ device, in particular by accepting a different measurement outcome, a different state will be heralded. This heralded state may be potentially useful for instance, we speculate that it would be possible to generate truncated cat-like states in this way.

Generalized scissors belong to a class of protocols known as tele-amplification \cite{neergaard-nielsen2013quantum}. Scissors are tele-amplification devices upon taking the amplitude of the entangled cat-state resource to zero (derivation presented in~\cref{sec:appendixscissors}).

%Since scissors herald states with a hard truncation in Fock space, but in general tele-amplification does not, we speculate that it may be beneficial for some protocols to use a general tele-amplification device rather than a scissor.

The laws of quantum physics puts absolute limits on the performance of probabilistic NLA~\cite{Pandey_2013}. A natural question to ask is how do the scissors compare against these ultimate bounds. Within the high-fidelity region NLAs have success probabilities that decrease exponentially with $N$ (the order of truncation), and this is an unavoidable consequence of attempting noiseless linear amplification~\cite{Pandey_2013}. Our scissors do not obtain the ultimate bound; one would require more complicated (probably highly nonlinear) devices, such as the proposal in Ref.~\cite{McMahon_2014}.% The fidelity and success probability together determine the overall performance of NLA devices.

%Our scissors do not obtain the ultimate bound on the success probability~\cite{Pandey_2013}, however this is the price to pay for such simple linear devices, employing just beamsplitters and photon detectors. 

In conclusion, we have introduced new quantum scissors which truncate and ideally amplify optical states using linear optical components. Compared to use of multiple scissors in parallel we found that the new scissors are more efficient for noiseless linear amplification and more practical for experimental implementation. This device may be scaled up to $2^S{-}1$ numbers of photons at the cost of a diminishing probability of success. We expect that our generalized scissors will in some situations 
improve the performance of existing experiments in quantum communication and make theorized protocols realizable in the near future. %Future work would be to examine how the generalized scissors compete in various protocols and scenarios. 

%In conclusion, we have presented a scissors device which is a linear optical realisation of non-deterministic noiseless linear amplification, 

%====================================================
%====================================================
%====================================================
%====================================================%====================================================
\begin{acknowledgements}

We thank Josephine Dias for valuable discussions during our investigation. This research was supported by the Australian Government Department of Defence and by the Australian Research Council (ARC) under the Centre of Excellence for Quantum Computation and Communication Technology.

\end{acknowledgements}

\appendix

\section{Derivations of generalized scissors\label{sec:appendixscissors}}

%%%%%%%%%%%%%%%%
\subsection{\label{sec:appendix1-scissor}Single photon quantum scissor}
%====================================================

\begin{figure}
        \includegraphics[width=\linewidth]{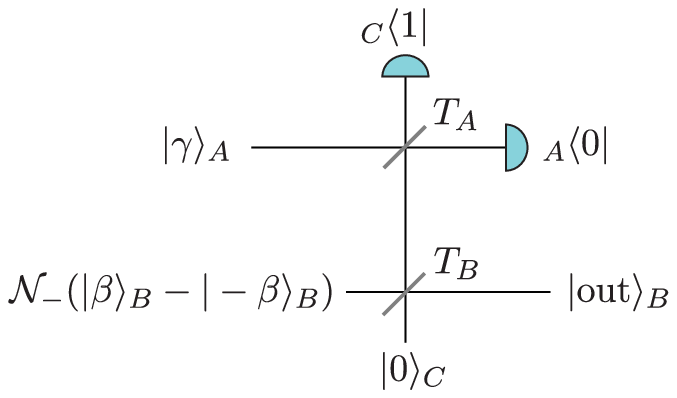}
          \caption{$\twocatrelay$ with two-component cat state as a resource and two-port detection scheme~\cite{neergaard-nielsen2013quantum}.}
          \label{fig:2cat_scissor}
          \end{figure}

We first reconsider the modified single photon quantum scissor ($\oneinjectedscissor$), as discussed in the main paper. For simplicity, we will consider how an arbitrary coherent state $\ket{\gamma}$ transforms. Rather than solving the problem for the $\oneinjectedscissor$ consisting of a single injected photon, it is convenient and intuitive to consider the tele-amplification procedure from~\cite{neergaard-nielsen2013quantum}, where a two-component cat state is used as the resource (replacing the single photon resource). This set-up is shown in \cref{fig:2cat_scissor} and we will refer to it as $\twocatrelay$. In the limit that $\alpha\to0$, the $\oneinjectedscissor$ is equivalent to $\twocatrelay$ and we show this for coherent states in what follows.

Let's assume that the initial input state of mode A is a coherent state $\ket{\gamma}_A$. The resource state of mode B is an odd (minus) 2-cat state, and the initial state of mode $C$ is vacuum. The three-mode state after the action of the two beamsplitters $\hat{V}_B$ and $\hat{V}_A$, with transmissivity $T_B$ and $T_A$ respectively, is
\begin{widetext}
\begin{equation}
\begin{split}
\ket{\psi}_{ABC} & = \mathcal{N}_{-} \hat{V}_A \hat{V}_B \ket{\gamma}_A ( \ket{\beta}_B{-}\ket{{-}\beta}_B )  \ket{0}_C\\
%\begin{multline}
& = \mathcal{N}_{-}  \ket{\sqrt{T_A} \gamma - \sqrt{(1-T_A)(1-T_B)} \beta}_A \ket{\sqrt{T_B}\beta}_B \ket{{-}\sqrt{1-T_A} \gamma - \sqrt{T_A(1-T_B)}\beta}_C\\
& - \mathcal{N}_{-} \ket{\sqrt{T_A} \gamma + \sqrt{(1-T_A)(1-T_B)} \beta}_A \ket{{-}\sqrt{T_B}\beta}_B \ket{{-}\sqrt{1-T_A} \gamma + \sqrt{T_A(1-T_B)}\beta}_C,
%\end{multline}
\end{split}
\end{equation}
\end{widetext}
where $\mathcal{N}_{-}=[2(1{-}e^{-2|\beta|^2})]^{-1/2}$ is the normalization constant for the cat state resource. To simplify the calculation, we introduce a new complex variable $\alpha$, and impose the condition $\beta = \alpha\sqrt{T_A/[(1-T_A)(1-T_B)]}$. To simplify things further, we also set $T_A = 0.5$ (but it may be advantageous to alter the transmissivity of beamsplitter $A$ in other scenarios). Single photon detection is performed on modes $A$ and $C$. Conditioned on no photon detection at $A$ and single photon detection at $C$, the final state of mode $B$ is
\begin{align}
{}_C\bra{1}{}_A\braket{0}{\psi}_{ABC}&= \mathcal{N} ( (\alpha+\gamma) \ket{g\alpha} + (\alpha-\gamma) \ket{{-}g\alpha}),
\end{align}
where $g = \sqrt{{T_A T_B}/{(1-T_A)(1-T_B)}}$, and $\mathcal{N}$ is given by
\begin{align}
\mathcal{N} &= \frac{- \mathcal{N_{-}} e^{-(|\gamma|^2+|\alpha|^2)/2}}{\sqrt{2}}.
\end{align}
We expand in the Fock basis to first order in $\alpha$ and we get
\begin{align}
&= \mathcal{N} e^{-|g\alpha|^2/2}  ( (\alpha{+}\gamma) (\ket{0} {+} g\alpha \ket{1}) {+} (\alpha-\gamma) (\ket{0} {-} g\alpha \ket{1})\\
&= \mathcal{N} e^{-|g\alpha|^2/2}  ( 2\alpha \ket{0} + 2g\alpha\gamma \ket{1})\\
&= \mathcal{N} \times  2\alpha e^{-|g\alpha|^2/2}  (  \ket{0} + g\gamma \ket{1}).
\end{align}
Expanding and throwing away higher order $\alpha$ we find that $\mathcal{N}_{-}=1/(2|\beta|)$, also removing the global phase minus sign, we have
\begin{align}
&=  \sqrt{\frac{1-T_B}{2}} e^{-|\gamma|^2/2}  (  \ket{0} + g\gamma \ket{1})\\
&=  \sqrt{\frac{1}{2(g^2+1)}} e^{-|\gamma|^2/2} (  \ket{0} + g\gamma \ket{1}),
\end{align}
which is the known result for a $\oneinjectedscissor$. The set-up also works for the opposite measurement outcome: no photon detection at $C$ and single photon detection at $A$, however there is a passive phase flip. The success probability is doubled since we keep both measurement outcomes: ${}_C\bra{1}{}_A\bra{0}$ and ${}_C\bra{0}{}_A\bra{1}$.

%%%%%%%%%%%%%%%%
\subsection{\label{sec:appendix3-scissor}Three photon quantum scissor}
%%%%%%%%%%%%%%%%

\begin{figure}
        \includegraphics[width=\linewidth]{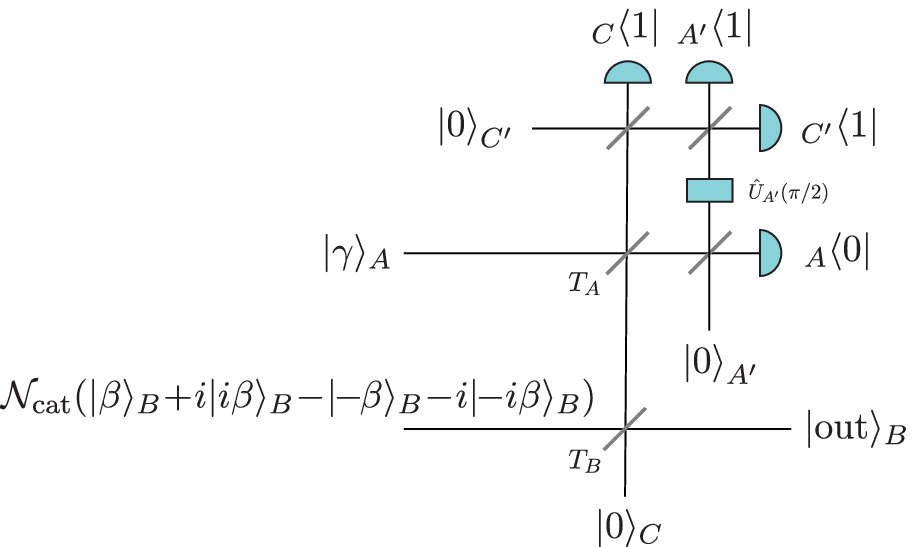}
          \caption{$\fourcatrelay$ with four-component cat state as a resource and four-port detection scheme~\cite{neergaard-nielsen2013quantum}.}
          \label{fig:4cat_scissor}
\end{figure}

Tele-amplification can be scaled up: cat states with more components are used as the resource, the detection procedure detects multiple single photons, and the device can tele-amplify coherent states from a larger alphabet. The next simplest case is $\fourcatrelay$ and is shown in \cref{fig:4cat_scissor}. In the limit that $\alpha\to0$, $\fourcatrelay$ is equivalent to a $\threeinjectedscissor$. In the following, we show how the $\threeinjectedscissor$ transforms coherent states by again considering tele-amplification.

To simplify the calculation, we introduce $\alpha$ and replace $\beta {=} \alpha\sqrt{T_A/[(1-T_A)(1-T_B)]}$. The gain is $g{=} \sqrt{{T_A T_B}/{(1-T_A)(1-T_B)}}$, and we set $T_A{=}0.5$ also at this stage. The entire state after propagating through the circuit, but before the measurement, is given by
\begin{multline}
\ket{\psi}_{BACA'C'} = \mathcal{N}_\text{cat} \sum_{k=0}^3 i^k \ket{g \alpha i^k}_B \otimes \ket{\frac{\gamma - \alpha i^k}{2}}_A \otimes \ket{\frac{-(\gamma + \alpha i^k)}{2}}_C \\ \otimes \ket{\frac{\gamma(1-i)+\alpha(i^k+i^{k+1})}{2\sqrt{2}}}_{A'} \otimes \ket{\frac{\gamma(1+i)+\alpha(i^k-i^{k+1})}{2\sqrt{2}}}_{C'},
\end{multline}
where $\mathcal{N}_\text{cat} = 1/\sqrt{8 e^{-\beta^2}(\sinh{\beta^2}-\sin{\beta^2})}$ is the normalization factor from the four-lobed cat state.

After the measurement ${}_C\bra{1}{}_{A'}\bra{1}{}_{C'}\bra{1}{}_A\bra{0}$ we get
\begin{multline}
{}_C\bra{1}{}_{A'}\bra{1}{}_{C'}\bra{1}{}_A\braket{0}{\psi}_{BACA'C'} \\ = \mathcal{N}_\text{cat} \sum_{k=0}^3 i^k   e^{-\smallfrac{|A|^2}{2}} e^{-\smallfrac{|A'|^2}{2}} e^{-\smallfrac{|C|^2}{2}} e^{-\smallfrac{|C'|^2}{2}}  A' C C' \ket{g \alpha i^k}_B,
\end{multline}
where
\begin{align}
A &= \frac{\gamma - \alpha i^k}{2}\\
A' &= \frac{\gamma(1-i)+\alpha(i^k+i^{k+1})}{2\sqrt{2}}\\
C &= \frac{-(\gamma + \alpha i^k)}{2}\\
C' &= \frac{\gamma(1+i)+\alpha(i^k-i^{k+1})}{2\sqrt{2}}.
\end{align}
Simplifying, we get
\begin{multline}
{}_C\bra{1}{}_{A'}\bra{1}{}_{C'}\bra{1}{}_A\braket{0}{\psi}_{BACA'C'} = \mathcal{N}_\text{cat} \sum_{k=0}^3 i^k  e^{-(|\gamma|^2+|\alpha|^2)/2} \\ \times \frac{1}{8} (\alpha i^k - \gamma i)(\gamma i +\alpha i^k)(\gamma + \alpha i^k) \ket{g \alpha i^k}_B.\label{eq:cat4relayder}
\end{multline}
We expand in the Fock basis to first order in $\alpha$
\begin{align}
&= \mathcal{N}_\text{cat} e^{-(|\gamma|^2+|\alpha|^2)/2}\frac{1}{2} \alpha^3 e^{-|g\alpha|^2/2}  \sum_{n=0}^3 \frac{g^n \gamma^n}{\sqrt{n!}} \ket{n}.
\end{align}
For small $\alpha$ note that $\mathcal{N}_\text{cat} = \sqrt{3}/(2\sqrt{2}\beta^3)$. Therefore, we have
\begin{align}
&= \frac{\sqrt{6}}{8 } \frac{\alpha^3}{\beta^3} e^{-(|\gamma|^2+|\alpha|^2)/2}  e^{-|g\alpha|^2/2}  \sum_{n=0}^3 \frac{g^n \gamma^n}{\sqrt{n!}} \ket{n}.
\end{align}
Recall that $\alpha/\beta = (1/(g^2+1))^{1/2}$, and throwing away all left-over high order terms in $\alpha$ we get
\begin{align}
&= \frac{\sqrt{6}}{8} \left(\frac{1}{g^2+1}\right)^\frac{3}{2} e^{-|\gamma|^2/2} \sum_{n=0}^3 \frac{g^n \gamma^n}{\sqrt{n!}} \ket{n}.
\end{align}
In the limit that $\alpha\to0$, the $\fourcatrelay$ is a generalized quantum scissor ideal NLA, truncated at Fock number 3. The generalized scissors truncate and amplify arbitrary input states (not only coherent states as shown above).

The different click patterns herald passive phase shifts on the state of 0, $\pi/2$, $\pi$, or $3\pi/2$. It can be corrected for by simply delaying the beam i.e. linear phase shift operation (or for some applications the phase shift may simply be tracked in software). Explicitly, depending on the result at the four ports $(A, A’, C, C’)$, the transformation is given by:
\begin{align}
\ket{\psi}&\to \hat{T}_\text{3}\ket{\psi}\text{, for }(0,1,1,1)\\
\ket{\psi}&\to e^{\smallfrac{\pi}{2}\hat{a}^\dagger\hat{a}}\hat{T}_\text{3}\ket{\psi}\text{, for }(1,0,1,1)\\
\ket{\psi}&\to e^{\pi\hat{a}^\dagger\hat{a}}\hat{T}_\text{3}\ket{\psi}\text{, for }(1,1,0,1)\\
\ket{\psi}&\to e^{\smallfrac{3\pi}{2}\hat{a}^\dagger\hat{a}}\hat{T}_\text{3}\ket{\psi}\text{, for }(1,1,1,0)
\end{align}
where $\hat{T}_\text{3}$ is the $\threeinjectedscissor$ transformation.

Cat states with even more lobes may be used which is equivalent to injecting more photons into the scissor (and checking for more photons). The next order device we present is a seventh order device, see \cref{sec:appendix7-scissor}. In the limit of a large number of injected photons, generalized scissors perform as an ideal NLA without truncation, but with a vanishing success probability.% But first, in \cref{sec:appendix2-scissor}, we wonder if it is possible to create a working $\twoinjectedscissor$.

%%%%%%%%%%%%%%%%
\subsection{\label{sec:appendix2-scissor}Two photon quantum scissor}
%%%%%%%%%%%%%%%%

%Creating two photons is experimentally less challenging than creating three photons. We would like to generalize the concept of quantum scissors to include scissors with an even number of injected photons. There are two options for designing a $\twoinjectedscissor$. Should it be a $\threecatrelay$ or a $\fourcatrelay$? 

We have introduced a three photon scissor. Is it possible to design a working two photon scissor ($\twoinjectedscissor$) which truncates and amplifies to second order? Recall that energy conservation requires that one of the output modes of a $50{:}50$ beamsplitter suffers a $\pi/2$ phase shift. It has been noted that phase conditions of $2\pi/3$ would be required to enable a working $\twoinjectedscissor$~\cite{jeffers2010nondeterministic} and the only way that this is possible would be to use a beamtritter, or lossy beamsplitters. On the other hand, the three photon scissor requires $\pi/2$ phase shifts and so it is possible with a simple circuit consisting of ordinary beamsplitters.

Therefore, we find that the scissors seem to generalize only to $2^S{-}1$. This is not surprising since it has been noted that teleportation protocols often rely on detecting odd numbers of photons~\cite{enk2003odd}.

That being said, we were in fact successful in creating a $\twoinjectedscissor$, but only for specific inputs such as coherent states $\ket{\gamma}$. Let us consider again $\fourcatrelay$. It requires three single photon detections. This device actually works with other Fock-shifted 4-cat resources, not just the ``correct'' cat state which is Fock state $|3\rangle$ in the limit of zero amplitude. We can instead use the cat state which goes to two photons $\ket{2}$ for small amplitude and create a working $\twoinjectedscissor$. Repeating the calculation in section \cref{sec:appendix3-scissor} but with a different cat state resource, the 4-cat state which goes to two photon numbers, and then taking $\alpha\to0$ we create a working $\twoinjectedscissor$ based on $\fourcatrelay$. However, it only works for specific inputs such as coherent states. The $\twoinjectedscissor$ is the same as the $\threeinjectedscissor$ except with two photons entering the device instead of three. The measurement procedure is the same (measure three single photon clicks, and vacuum at the other).

The $\twoinjectedscissor$ is identical to the $\threeinjectedscissor$ shown in the main text, except injecting a two photon state $\ket{2}$ instead of a three photon state $\ket{3}$. The $\twoinjectedscissor$ transforms coherent states as
\begin{multline}
\hat{T}_{\twoinjectedscissor}\ket{\gamma} = \gamma \frac{\sqrt{2}}{8} \left(\frac{1}{g^2+1}\right) e^{-|\gamma|^2/2} \\ \times  (  |0\rangle + g \gamma |1\rangle + \frac{g^2 \gamma^2}{\sqrt{2}} |2\rangle),\label{eq:2scissor}
\end{multline}
thus acting as an ideal NLA for coherent state inputs of sufficiently small amplitude.

The success probability of the $\twoinjectedscissor$ goes as $\gamma^2/(g^2+1)^2$, so in general in the regime of high fidelity, the success probability of the $\threeinjectedscissor$ is higher than for the $\twoinjectedscissor$. Since the success probability is higher, the fidelity better, and there is no restriction on the input state, it is better to use the $\threeinjectedscissor$.

This $\twoinjectedscissor$ result has surprisingly advantageous consequences for the $\threeinjectedscissor$ performance. For coherent state inputs, the $\threeinjectedscissor$ device allows for loss in the resource mode before the beamsplitter. What we mean by this is that if the resource mode $\ket{3}$ of the $\threeinjectedscissor$ loses a photon such that $\ket{2}$ enters the device, the device will still successfully truncate and amplify the coherent state input, although now truncating at $\ket{2}$. Thus, the scissor is resilient to loss in this mode.

The reason it works for coherent states is that $\gamma$ is pulled out of all terms and can be absorbed into the normalization constant. Seemingly magically, the factors $g$ and $1/\sqrt{n!}$ are in the correct place. This is because the factors $g$ and $1/\sqrt{n!}$ come from cat components $\ket{g\alpha i^k}_B$ and are not scrambled, whereas $\gamma$ comes from the input state and is scrambled. For coherent states, this scrambling is okay. For other input states, the coefficients on all the Fock states in superposition will be scrambled. 

This $\twoinjectedscissor$ may still be useful for entanglement distillation but it certainly does fail to transform an EPR state in the ideal truncation and amplification way. Specifically, applying the $\twoinjectedscissor$ to one arm of an EPR state results in a state that looks like $\ket{10}+\ket{21}+\ket{32}$ which could potentially be useful since it is still an entangled state. The scissor can be thought of as a photon shifter in this sense.

%%%%%%%%%%%%%%%%
\subsection{\label{sec:appendix7-scissor}Seven photon quantum scissor}
%%%%%%%%%%%%%%%%

\begin{figure}
      \makebox[\linewidth][c]{\includegraphics[width=\linewidth]{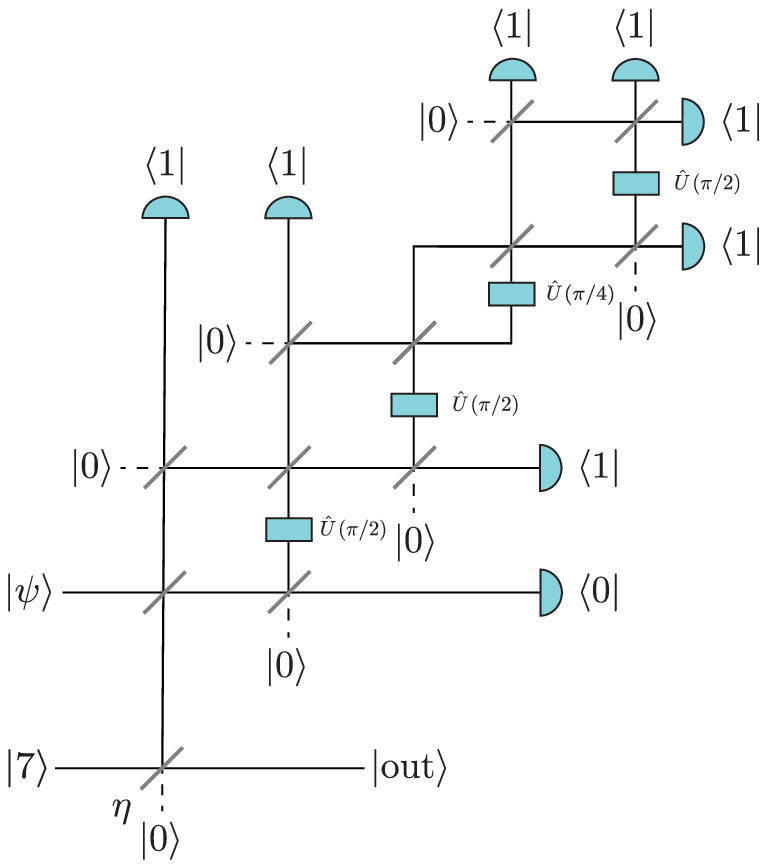}}
          \caption{Our generalized seven photon quantum scissor ($\seveninjectedscissor$) with injection of a seven photon state and eight-port single photon measurement scheme. The measurement process requires seven single photon detections and one no photon detection. The upper beamsplitters are all set to have transmissivity $1/2$. The lower beamsplitter has transmissivity $\eta$ which sets the gain of the NLA. The beamsplitters and phase shifts are arranged such that the input state is teleported, truncated and amplified to the output. Replacing $|7\rangle$ with an 8-lobed cat state, then this device tele-amplifies an arbitrary superposition of eight coherent states on a ring in phase space~\cite{neergaard-nielsen2013quantum}.}
          \label{fig:7scissor}
\end{figure}

Our scissors generalize to higher order, specifically, we speculate at least to $2^S{-}1$. The simplest scissor is the $\oneinjectedscissor$ for $S{=}1$. And then the $\threeinjectedscissor$ with $S{=}2$. The next device is $S{=}3$ which is a seven photon scissor ($\seveninjectedscissor$). This scissor is shown in~\cref{fig:7scissor}.

The detection scheme is intuitive if you consider the device from a tele-amplification point of view for 8-lobe cat states. Note that the $\pi/4$ phase shift and surrounding beamsplitter interferometer is can be re-written in a more practically implementable way by removing one of the beamsplitters and solving for the correct phase shift and transmissivity, however, we present the device as shown because it is more intuitive to see how the device works. The lower left part of the detection scheme is reminiscent of our $\threeinjectedscissor$ (which is itself a simple extension of the $\oneinjectedscissor$). After the $\pi/4$ phase shift (required since 8 lobe cat states consist of superpositions of coherent states with phase multiples of $\pi/4$ around the circle in phase space), the upper right four photon detections complete the ideal tele-amplification up to seventh order.

%%%%%%%%%%%%%%%%
\section{\label{sec:appendixPsuc}Success probabilities}
%%%%%%%%%%%%%%%%

In this section we write down the success probabilities. The probability of success is given by the squared norm of the unnormalized state which depends on the input state. 

For a single photon quantum scissor ($\oneinjectedscissor$), the success probability is
\begin{align}
P_{\oneinjectedscissor} &=  2 \times \frac{1}{2(g^2+1)} (|c_0|^2+|g c_1|^2).
\end{align}
The squared norm has been multiplied by 2 because we assume that the phase shift on the single photon component given by the reverse click pattern can be corrected with a phase-shifter.

For a $\threeinjectedscissor$, the success probability is
\begin{multline}\label{eq:PROB}
P_{\threeinjectedscissor} = 4 \times \frac{3}{32} \left(\frac{1}{g^2+1}\right)^{3} \\ \times (|c_0|^2+|g c_1|^2 +|g^2 c_2|^2+|g^3 c_3|^2),
\end{multline}
where we multiply by a factor of 4 which accounts for the other three measurement outcomes, which lead to a heralded phase shift of the state which can be passively corrected for with a phase-shifter.

The success probability for $N$ $\oneinjectedscissors$ in parallel is
\begin{align}
P_{\Nscissors} &=  \left(\frac{1}{g^2+1}\right)^{N}  \left( \sum_{n=0}^{N} \left| \frac{N!}{(N-n)!N^n} g^n c_n \right|^2 \right),
\end{align}
assuming again that phase flips in each $\oneinjectedscissor$ are corrected for by feeding forward. The total success probability decreases with $N$. For large $N$, the device approaches an ideal NLA and has a vanishing success probability.

\section{\label{sec:appendixmultiplescissors}NLA based on multiple single photon scissors in parallel}

The scheme for NLA from Ref.~\cite{ralph2009nondeterministic} performs the transformation
\begin{align}
\hat{T}_{\Nscissors} \ket{\psi} = \left(\frac{1}{g^2+1}\right)^\frac{N}{2} \sum_{n=0}^{N} \frac{N!}{(N-n)!N^n} g^n c_n \ket{n}.
\end{align}
NLA based on multiple single photon scissors perform distorted amplification for practical numbers of scissors $N$. It is assumed that for NLA based on multiple single photon scissors, phase flips on the one-photon component for each individual scissor are corrected by feeding forward (thus doubling the success probability of each scissor in the NLA), i.e., for $\twoscissors$ in parallel
\begin{align}
\hat{T}_{\twoscissors} \ket{\psi} &= \frac{1}{g^2+1}   (c_0 |0\rangle + g c_1 |1\rangle + \frac{1}{2} g^2 c_2 |2\rangle ),
\end{align}
\begin{align}
P_{\twoscissors} &=  \left(\frac{1}{g^2+1}\right)^{2} (|c_0|^2 + |g c_1|^2 + \frac{1}{4}|g^2 c_2|^2),
\end{align}
and for $\threescissors$
\begin{multline}\label{AAA}
\hat{T}_{\threescissors} \ket{\psi} = \left(\frac{1}{g^2+1}\right)^\frac{3}{2}  \\ \times (c_0 |0\rangle + g c_1 |1\rangle + \frac{2}{3} g^2 c_2 |2\rangle +  \frac{2}{9} g^3 c_3 |3\rangle ),
\end{multline}
\begin{multline}
P_{\threescissors} =  \left(\frac{1}{g^2+1}\right)^{3} \\ \times (|c_0|^2 + |g c_1|^2 + \frac{4}{9}|g^2 c_2|^2 + \frac{4}{81}|g^3 c_3|^2).
\end{multline}

\section{\label{sec:appendixEPR}Entanglement distillation}
In this section, we consider the action of scissors applied to one arm of an EPR state.

Consider transmitting the second mode of an EPR state through a pure loss channel of transmissivity $T$. The global state is given by
\begin{multline}
|\chi\rangle_{AA'E} \to  \sqrt{1-\chi^2} \sum_{n=0}^\infty \chi^n \sum_{k=0}^n \sqrt{{n \choose k}} \\ \times (1-T)^{k/2} T^{(n-k)/2}|n\rangle_{A} |n-k\rangle_{A'} |k\rangle_{E},
\end{multline}
and the trace is taken over mode E so the state is mixed. 
Following the pure loss channel by a $\oneinjectedscissor$ results in a state with density operator
\begin{align}
\rho_{AB} &= \sum_{k=0}^{\infty} |\psi_k \rangle \langle \psi_k |,
\end{align}
where 
\begin{multline}
|\psi_k \rangle = \sqrt{\frac{1-\chi^2}{2(g^2+1)}} (1-T)^\frac{k}{2} \\ \times [(-\chi)^k |k\rangle |0\rangle + \sqrt{T(k+1)} (-\chi)^{k+1} g |k+1\rangle |1\rangle].
\end{multline}
Similarly, following the pure loss channel by a $\threeinjectedscissor$ gives
\begin{align}
\rho_{AB} &= \sum_{k=0}^{\infty} |\psi_k \rangle \langle \psi_k |,
\end{align}
where
\begin{multline}
|\psi_k \rangle = \frac{\sqrt{6}}{8} \sqrt{\frac{1-\chi^2}{(g^2+1)^3} } \sum_{n=k}^{n+3} \chi^n g^{n-k} \sqrt{{n \choose k}} \\ \times (1-T)^{k/2} T^{(n-k)/2} \ket{n}\ket{n-k}.
\end{multline}

%%%%%%%%%%%%%%%%
\section{\label{sec:appendixRCI}RCI}
%%%%%%%%%%%%%%%%

The total RCI of a state $\rho_{AB}$ is defined as~\cite{PhysRevLett.102.210501}
\begin{align}
\text{RCI}(\rho_{AB}) &= H(A) - H(AB),
\end{align}
where $H(A)$ and $H(AB)$ are von Neumann entropies of $\rho_A = \text{Tr}_B(\rho_{AB})$ and $\rho_{AB}$ respectively. The von-Neumann entropy of density matrix $\rho$ is $-\text{Tr}(\rho \text{log}_2 \rho)$.

The Gaussian RCI is calculated from the covariance matrix of $\rho$:
\begin{align}
\text{Gaussian RCI} &= g(\vec{\nu}_A)-g(\vec{\nu}_{AB}),
\end{align}
where $g(x) = \frac{x+1}{2}\text{log}_2(\frac{x+1}{2}) - \frac{x-1}{2}\text{log}_2(\frac{x-1}{2})$, and $\vec{\nu}_A$ and $\vec{\nu}_{AB}$ are the symplectic eigenvalues of the covariance matrices of $\rho_A$ and $\rho_{AB}$ respectively.

%%%%%%%%%%%%%%%%
\section{\label{sec:appendixDetectors}Non-ideal single photon detectors, on-off detectors and loss at resource mode}
%%%%%%%%%%%%%%%%

Generalized scissors are robust to inefficient single photon detectors, only the success probability is affected. More experimentally feasible than single photon detectors are on-off detectors. The on-off measurement operators are
\begin{align}
\hat{\Pi}_\text{off} &=  \ketbra{0}{0},\\
\hat{\Pi}_\text{on} &=  \Id - \ketbra{0}{0}.
\end{align}

With on-off detection, \cref{eq:cat4relayder} for example needs correcting; the output state is no longer pure and is given by the density operator
\begin{align}
\rho_\text{out} = \sum_{n,m,l=1}^\infty \ket{\psi_{n,m,n}}\bra{\psi_{n,m,n}},
\end{align}
where
\begin{multline}
\ket{\psi_{n,m,n}} = \mathcal{N}_\text{cat} \sum_{k=0}^3 i^k   e^{-\smallfrac{|A|^2}{2}} e^{-\smallfrac{|A'|^2}{2}} e^{-\smallfrac{|C|^2}{2}} e^{-\smallfrac{|C'|^2}{2}} \\ \times \frac{{A'}^n}{\sqrt{n!}} \frac{{C{}}^m}{\sqrt{m!}} \frac{{C'}^l}{\sqrt{l!}} \ket{g \alpha i^k}_B.
\end{multline}

\begin{figure}
\centering
\includegraphics[width=\linewidth]{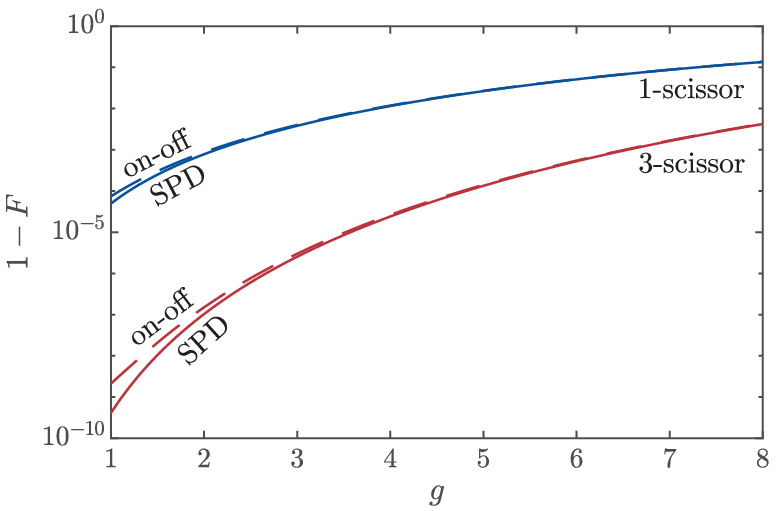}
\caption{a) Infidelity $1-F$ vs. the gain $g$ for the $\oneinjectedscissor$ and the $\threeinjectedscissor$ with perfect single photon detectors (SPD) or on-off detectors. The input state is a coherent state $\ket{\gamma}$ with magnitude $\gamma=0.1$. (There is very little increase in the probability of success for $\gamma = 0.1$ by using on-off detectors, thus, we have not included a similar plot showing the success probability.)\label{fig:onoff}}
\end{figure}

We plot the infidelity vs. gain in~\cref{fig:onoff} for single photon detectors (SPD) and on-off detectors, (assuming perfect efficiency for all devices). The input state is a coherent state with magnitude $\gamma = 0.1$. In the high-fidelity regime of the scissors (low amplitude input states), this correction is small and the scissors perform well with on-off detectors in place of SPDs.

%In summary, we have found that scissors are only a little negatively affected by non-perfect SPDs, or using on-off detectors in place of SPDs.

\begin{figure}
\centering
\includegraphics[width=0.95\linewidth]{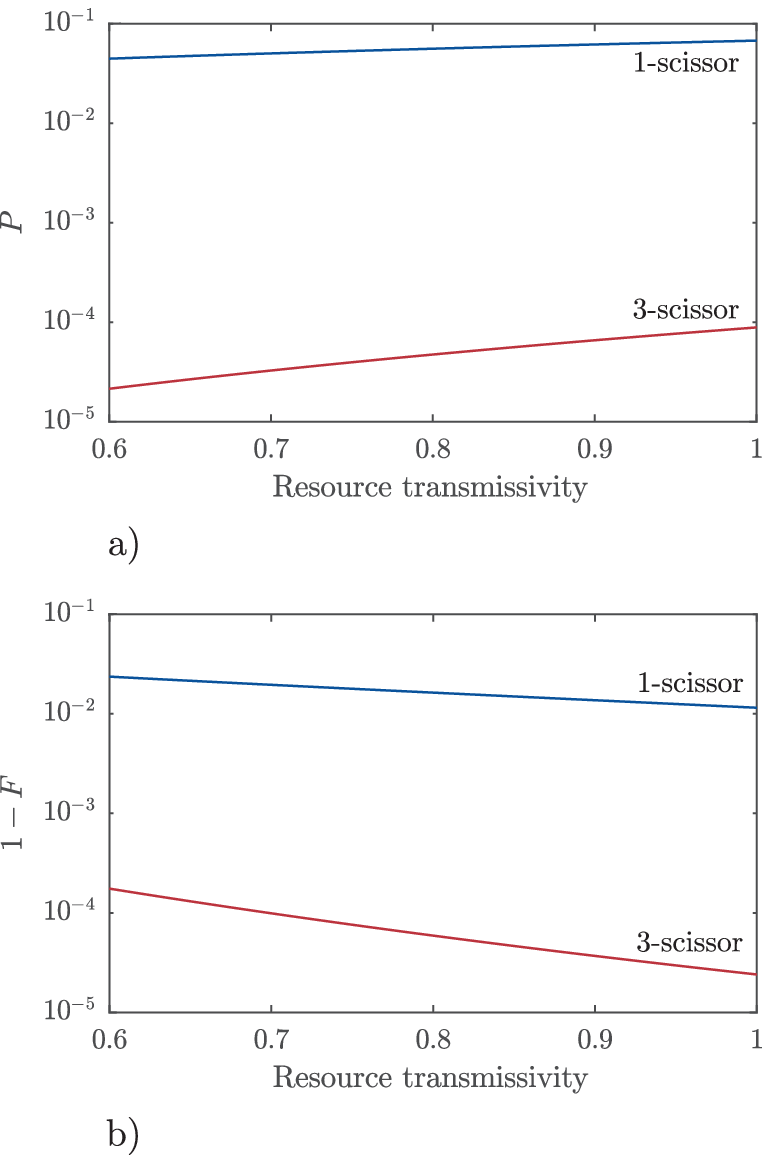}
\caption{a) Probability of success $P$ and b) infidelity $1-F$ vs. transmissivity $\tau_s$ of the resource mode for the $\oneinjectedscissor$ and the $\threeinjectedscissor$. The input state is a coherent state $\ket{\gamma}$ with magnitude $\gamma=0.1$. The gain is set to $g=4$.\label{fig:lossatresource}}
\end{figure}

Of greater concern for the practicality of generalized scissors is the efficiency of the resource Fock state. In~\cref{fig:lossatresource} we plot the success probability and infidelity of the output state with the target state $|g\gamma\rangle$ as a function of transmissivity $\tau_s$ of the resource mode. The input state is a coherent state $\gamma=0.1$ and the gain is fixed to $g=4$. For coherent state inputs, the scissors are surprisingly robust to resource inefficiencies. This can be explained by our $\twoinjectedscissor$ result, derived in~\cref{sec:appendix2-scissor}.

%z

\bibliography{bibfile.bib}
%Control: production of eprint (1) enabled

%Control: production of eprint (1) enabled

%====================================================
%====================================================

\end{document}